\documentclass[11pt,a4paper]{ivoa}
\newif\iftth
\iftth
\newcommand{\ivoaDocversion}{1.1}
\newcommand{\ivoaDocdate}{2017-12-15}
\newcommand{\ivoaDocdatecode}{20171215}
\newcommand{\ivoaDoctype}{NOTE}
\newcommand{\ivoaDocname}{cubeDM}

\input tthntbib.sty

\begin{html}
<meta http-equiv="Content-type" content="text/html;charset=UTF-8"/>
\end{html}

\definecolor{ivoacolor}{rgb}{0.0,0.318,0.612}

\renewcommand{\author}[2][0]{\def\@tmp{#1}
  \if 0\@tmp
	{\begin{html}<li class="author">\end{html}#2\begin{html}</li>\end{html}}\else
	{\begin{html}<li class="author"><a href="#1">\end{html}#2\begin{html}</a></li>\end{html}}\fi}
\renewcommand{\previousversion}[2][0]{\def\@tmp{#1}
  \if 0\@tmp
	{\begin{html}<li class="previousversion">#2</li>\end{html}}\else
	{\begin{html}<li class="previousversion">
	  <a href="#1">#2</a></li>\end{html}}\fi}
\renewcommand{\ivoagroup}[1]
  {\begin{html}<dd id="ivoagroup">#1</dd>\end{html}}
\renewcommand{\editor}[2][0]{\def\@tmp{#1}
  \if 0\@tmp
        {\begin{html}<li class="editor">\end{html}#2\begin{html}</li>\end{html}}\else
        {\begin{html}<li class="editor"><a href="#1">\end{html}#2\begin{html}</a></li>\end{html}}\fi}

\newcommand{\includeMeta}{%
   \ivoaDocversion\ivoaDoctype\ivoaDocname\ivoaDocdate}

\def\SVN$#1: #2 ${%
	#2}

\newenvironment{abstract}{%
  \includeMeta
  \begin{html}
    </div> <!-- titlepage -->
    <div id="abstract"><h2>Abstract</h2>
  \end{html}
  }{%
    \ivoaDoctype
    \tableofcontents
  }

\newcommand{\lstloadlanguages}[1]{}
\newcommand{\lstset}[1]{}

\newenvironment{lstlisting}[1][plain]
  {\tthverbatim{lstlisting}}
  {}

\renewcommand{\customcss}[1]{%
  \begin{html}<span class="customcss" ref="#1"/>\end{html}}

\newcommand{\harvarditem}[4][0]{%
  
  \if 0#1 \item[#2 (#3)]
  \else \item[#1 (#3)]\fi}
\newcommand{\harvardurl}[1]{\url{#1}}

\def\AtBeginDocument#1{\relax}
\def\pgfsyspdfmark#1#2#3{\relax}

\newbox\voidb@x
\def\@m{\relax}

\begin{html}
  <div id="titlepage">
\end{html}

\fi

\usepackage{listings}
\usepackage{todonotes}
\usepackage{url}
\usepackage[section]{placeins}

\usepackage[utf8]{inputenc}
\usepackage[export]{adjustbox}

\definecolor{texcolor}{rgb}{0.4,0.1,0.1}

\iftth
  
\fi

\iftth
 \newcommand{\comicstuff}[1]{
    \begin{html}<span class="comic">#1</span>\end{html}}
\else
  \newcommand{\comicstuff}[1]{(HTML exclusive material)}
\fi
\customcss{custom.css}

\title{Time Series Cube Data Model}

\ivoagroup{Time domain interest group}

\author[http://wiki.ivoa.net/twiki/bin/view/IVOA/JiriNadvornik]{Jiří Nádvorník}
\author[http://wiki.ivoa.net/twiki/bin/view/IVOA/PetrSkoda]{Petr Škoda}
\author[http://wiki.ivoa.net/twiki/bin/view/IVOA/DaveMorris]{Dave Morris}
\author{Pavel Tvrdík}

\editor{Jiří Nádvorník}

\begin{document}

\begin{abstract}
The purpose of this document is to create a data model and its serialization for expressing generic time series data. Already existing IVOA data models are reused as much as possible. The model is also made as generic as possible to be open to new extensions but at the same time closed for modifications. This enables maintaining interoperability throughout different versions of the data model.

We define the necessary building blocks for metadata discovery, serialization of time series data and understanding it by clients.

We present several categories of time series science cases with examples of implementation. We also take into account the most pressing topics for time series providers like tracking original images for every individual point of a light curve or time-derived axes like frequency for gravitational wave analysis.

The main motivation for the creation of a new model is to provide a unified time series data publishing standard - not only for light curves but also more generic time series data, e.g., radial velocity curves, power spectra, hardness ratio, provenance linkage, etc. 

The flexibility is the most crucial part of our model - we are not dependent on any physical domain or frame models. While images or spectra are already stable and standardized products, the time series related domains are still not completely evolved and new ones will likely emerge in near future. That is why we need to keep models like \textit{Time Series Cube DM} independent of any underlying physical models. In our opinion, this is the only correct and sustainable way for future development of IVOA standards.
\end{abstract}

\section*{Acknowledgements}
This work was supported by grant LD-15113 of Ministry of Education Youth and Sports of the Czech Republic and by grant No. SGS16/123/OHK3/1T/18 from Czech Technical University in Prague.

\newpage

\section {Term Definition}

\begin{itemize}
\item {Data cube} - When we are speaking about a data cube, we have in mind the pure database point of view. In astronomy a term data cube is often implicitly understood as an image cube or a spectral cube and we do not want to be restricted only to these domains. In this document, a data cube is much closer to an OLAP data cube (\cite{kimball_olap}) (while again not restricting us to business-related domains). This data cube is usually implemented in the database as a star-schema with its dimensions being our axis domains (\cite{kimball_cube}).
\item {Time Series Cube DM} - The \textit{Time Series Cube data model (DM)} defines data structures for holding metadata and data of a time series data cube.
\item {Axis domain} - By axis domain we mean the physical interpretation of an axis of the data cube. This is usually axis metadata, e.g., spectral limits of that axis, resolution, etc. From the IT perspective, while \textit{Time Series Cube DM} holds the \textit{data}, axis domain model stores the relationships and context which enables us to get \textit{information} from the data.
\item {Dataset} - We mean an IVOA Dataset as specified in \textit{Observation Data Model} (\cite{std:OBSCORE}), also known as the \textit{ObsCore DM}, i.e. a collection in which can be stored any kind of observation. As such, a dataset is holding information common for every element of its collection, e.g., facility, instrument or creator.
\end{itemize}

\section{Introduction}	

This is a working draft of a new approach for representing generic time series data. It was designed to be modular and extensible, new changes will extend the model, not modify it. Extending a model means reusing what was in the extended model, while adding new structures to it. That said, any part of this design is open to discussion. If you have any ideas how it can be improved, please contact us on Time Domain Interest Group mailing list \href{mailto:voevent@ivoa.net}{voevent@ivoa.net}.

We can incorporate any kind of axis in the data cube, where all other up-to-now approaches tried to embed domain-specific metadata inside the model. They described specialized axes relevant only for light curves - SimpleTimeSeries (\cite{std:SimpleTimeSeries}) and SSAP-based approaches (\cite{std:SSAP}). For these approaches, making any enhancement to the model required re-work of what was already in the model, not an extension. 

The description of axis domain, i.e. the metadata which enables us to extract information and knowledge from the data, is kept outside of this model. The relationships towards other models can be found in Chap. \ref{chap:depend_model}.

Compared to previous approaches, we are only referencing the physical domain specific metadata, instead of incorporating them inside the \textit{Time Series Cube DM}. The detailed description of the \textit{Time Series Cube DM} can be found in Chap. \ref{chap:model}

\begin{figure}[h!]
\includegraphics[width=1\textwidth]{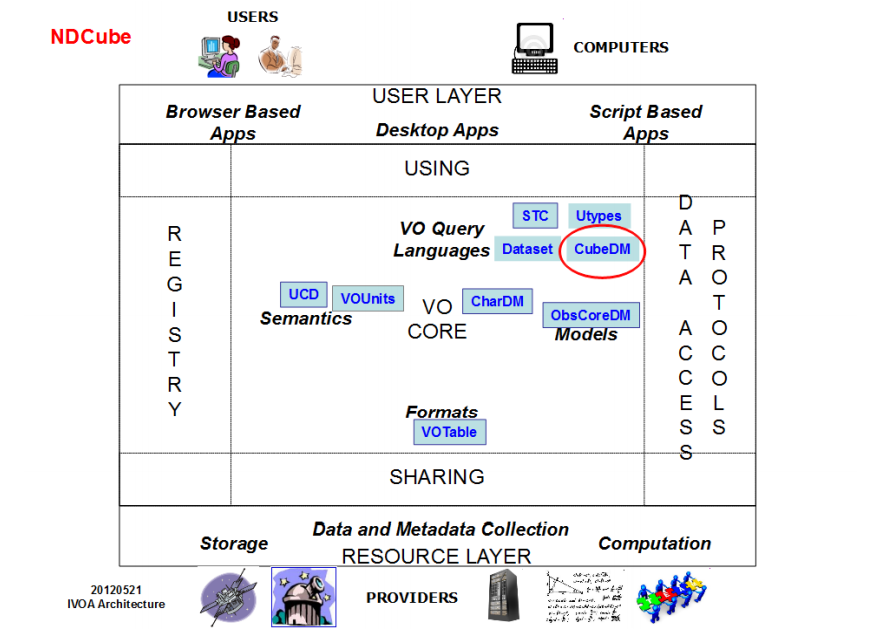}
\centering
\caption{IVOA architecture}
\label{fig:ivoa_model}
\end{figure}

The position of \textit{Time Series Cube DM} inside IVOA Architecture (\cite{note:VOARCH}) can be seen on Fig. \ref{fig:ivoa_model}.

\section{Dependent data models}
 \label{chap:depend_model}

In this chapter we describe \textit{Time Series Cube DM} position and relationships to other models. 

\begin{figure}[h!]
\includegraphics[width=0.5\textwidth]{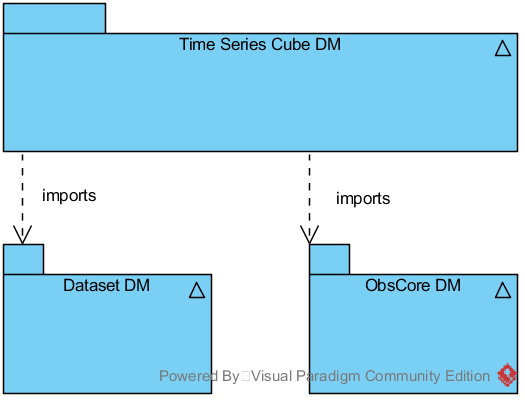}
\centering
\caption{This UML describes models that we import AS-IS, without any modifications to them or extending them.}
\label{fig:importUML}
\end{figure}

On Fig. \ref{fig:importUML} we can see other models that we import without modifications, i.e. copy them inside our model, and without extending them.

\begin{figure}[h!]
\includegraphics[width=0.5\textwidth]{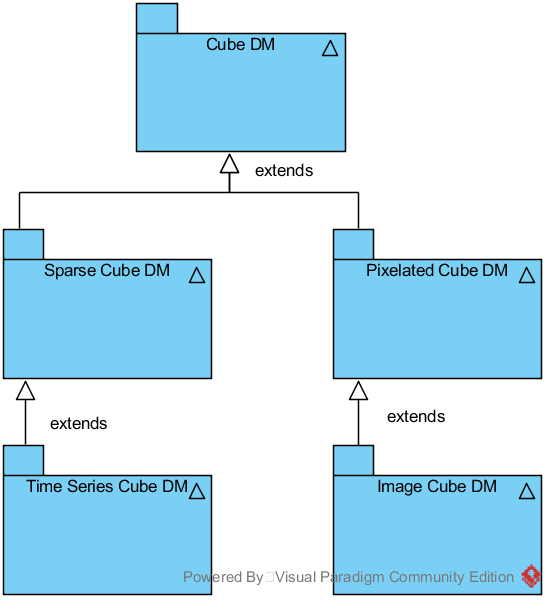}
\centering
\caption{UML model specifying the position of the \textit{Time Series Cube DM} compared to the \textit{Cube DM} and the \textit{ImageCube DM}.}
\label{fig:extendUML}
\end{figure}

On Fig. \ref{fig:extendUML} we show the \textit{Time Series Cube DM} relationships to other data cube models.

\subsection{Imported Data models}
We import these data models without modification. We basically copy the model definition into \textit{Time Series Cube DM}, making us completely dependent on what is defined within these - \textit{Time Series Cube DM} wouldn't work without them.

However, because we use them as black boxes, it does not matter if these data models change - \textit{Time Series Cube DM} has no common parts shared with them.

\subsubsection {Dataset DM}
The \textit{Dataset DM} is used for describing common information for the whole set of time series cubes. The individual cubes, i.e. elements of the dataset collection described by \textit{Dataset DM}, are then described within the \textit{Time Series Cube DM}.

\subsubsection{ObsCore DM}
 \label{chap:discovery}
\textit{ObsCore DM} in combination with TAP (\cite{std:OBSCORE}) is used for discovery of the time series cube metadata and we use it without modification. The \textit{ObsCore DM} has already support for time series data with \texttt{dataProductType=timeseries} parameter. We are now working with a reference implementation, using GAVO DaCHS for server side and modified Splat-VO (\cite{paper:splat}) for consumption of this ObsCore TAP service and discovering time series. More about discovery can be found in Chap. \ref{chap:discovery}.

\subsubsection{VO-DML}
The \textit{Time Series Cube DM} uses the VO-DML (\cite{std:VODML}) for annotating parts of the model and referencing individual axis domain models from inside the time series cube. The serialization of \textit{Time Series Cube DM} model into VO-DML is not the purpose of this document. 

We use VO-DML because it is the only reasonable way how to reference an actual model, not only entities already defined by that model. This is useful for referencing axis domains from within the axis definitions inside the \textit{Time Series Cube DM}.

We are not importing the whole VO-DML, the only purpose it serves us is brining a simple way to reference other models as a whole, not only their structures.

\subsection{Extended Data models}
The following models are either base classes or siblings with relationships to the \textit{Time Series Cube DM}. We reuse parts of them or reference them and in some cases, we override what has been defined in them if it was not generic enough for the purposes of \textit{Time Series Cube DM}.

\subsubsection {Cube DM}
The \textit{Cube DM} describes generic data cubes and is defined in the \textit{N-Dimensional Cube/Image Model} (\cite{std:NDCube}) document. This document is also describing the \textit{Sparse Cube DM} and the \textit{Pixelated Cube DM}, however, it is not describing \textit{Image Cube DM} (as could be erroneously understood from the title). We will refer to this standard as \textit{N-Dimensional Cube DM}.

\subsubsection {Pixelated Cube DM}
This data model describes an abstraction of pixelated data cubes, e.g., image cubes. Essentially this is still the same data cube as the generic one, we only put restrictions on the main axes. The pixelated cube must have pixelated axes, i.e., regular intervals between coordinates.

\subsubsection {Image DM}
The \textit{Image Cube DM} is currently being worked upon in the \textit{IVOA Image DM} (\cite{std:imageDM}) working draft but is not being aligned completely with the \textit{Cube DM} effort. It sis an extension of \textit{Pixelated Cube DM} and the image axes have naturally regular intervals between their coordinates.

\subsubsection {Sparse Cube DM}
This data model extends the \textit{Cube DM} while adding restrictions on the main axes. The main axes must hold sparse data, i.e., we can have irregular intervals between the coordinates.

\subsubsection{Time Series Cube DM}
The \textit {Time Series Cube DM} is describing time series data cubes, which are by definition sparse. We describe them in more detail in Chap. \ref{chap:model}.

\section{Time Series Cube DM}
\label{chap:model}

\begin{figure}[h!]
\includegraphics[width=1\textwidth]{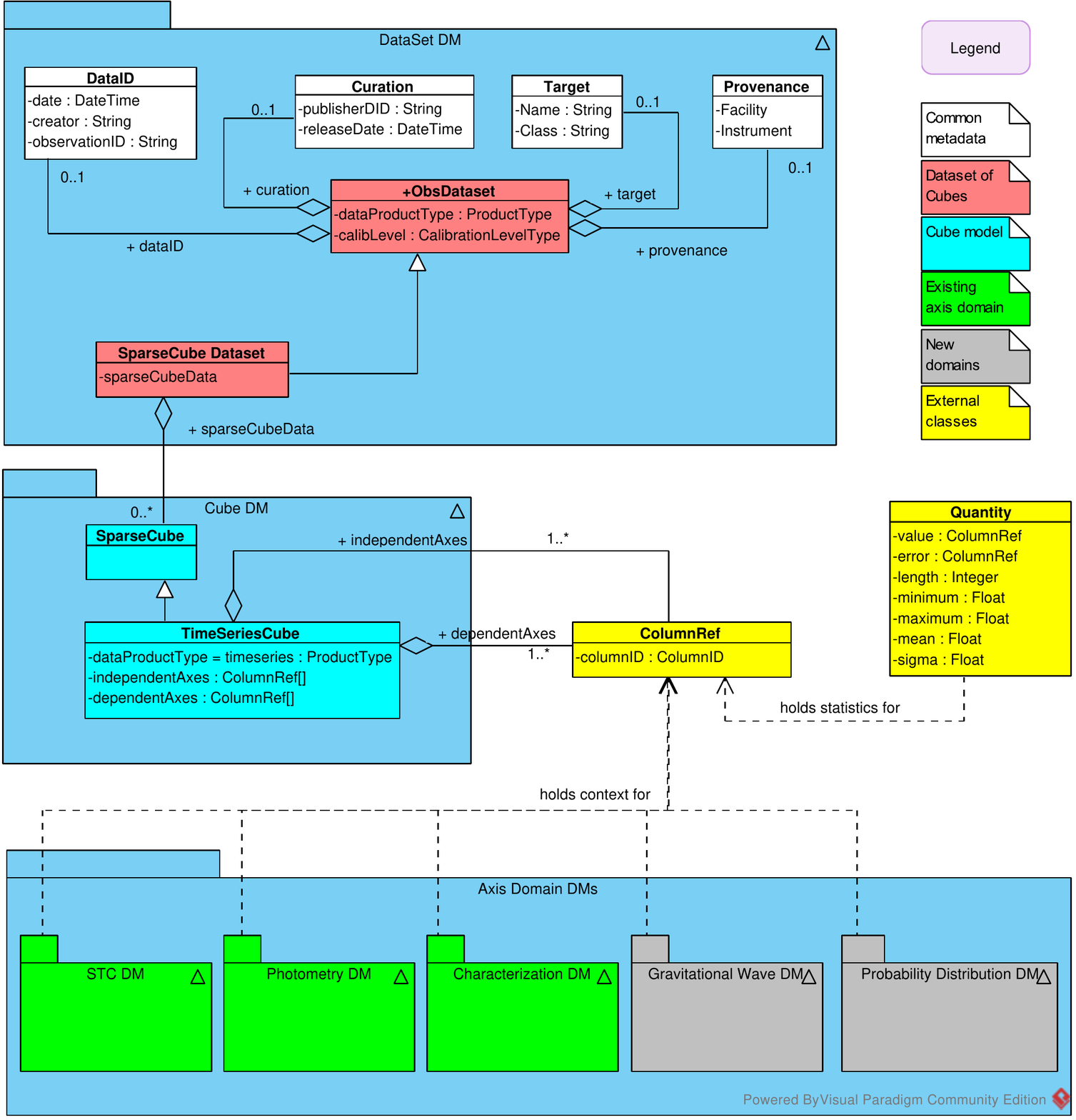}
\centering
\caption{The UML diagram of Time Series Cube DM ecosystem. The \textit{DataSet DM} classes (red) collect sparse cubes and time series cubes (cyan). The \texttt{Axis Domain DMs} referenced by the \texttt{CubeAxis} class are split between already existing (green) and potentially coming (grey) models.}
\label{fig:model}
\end{figure}

In this chapter we work with the time series data cubes defined within the \textit{Time Series Cube DM}. These data cubes are instances of the \texttt{Time Series Cube} class seen on Fig. \ref{fig:model}.

Note that the purpose of the \textit{Time Series Cube DM} metadata is to enlist all the axes of the cube and provide linkage towards their values and errors, not to completely describe axis domains. 

Also, after intensive discussions, we decided to move statistical metadata about the axis of our data cubes to be moved into a really small external model \textit{Quantity DM} because this will be useful not only for cubes but any numerical data in general.

The axis domain descriptions will be held within other specialized data models, e.g., spatial and time domain in STC (\cite{std:STC}), spectral domain in Photometry DM (\cite{std:PHOTDM}), etc. These external models will only be referenced by the \textit{Time Series Cube DM}. This is where we differ from the original \textit{Cube DM} that already focuses on supporting SIAv2 (\cite{std:SIAv2}) image-based protocol and we do not want to restrict ourselves only to pixelated types of data cubes. 

Withdrawing axis domain definitions as a dependency from the \textit{Time Series Cube DM} helps us in the future when new data models will be created, e.g., for probability distributions or gravitational wave domains. If such a new data model appears, we can just add it as a reference to axis metadata within the data cube and the \textit{Time Series Cube DM} does not need to be changed at all. It also solves the main drawback of previous approaches and that is extensibility - SSAP or SimpleTimeSeries standards would need to be changed if a new domain was to be supported. Moreover, such a domain would not be logically part of spectral or time aspects of those models and therefore it does not belong there.

The only difference between axis coordinate and the value for that coordinate is that the axis coordinates are independent while the data values are fully functionally dependent on these coordinates.

Therefore, we can afford (from the data point of view) to treat them equally, which is one of the biggest advantages of a data cube. We can browse the data cube by filtering the axis coordinates and the data values in the same way. This enables a client for example to slice the time axis of a light curve (coordinates) within an interval and the magnitude axis (values) within another interval using the same algorithm.

The time series data cube serialized accordingly to the \textit{Time Series Cube DM} describes only the data, not the physical interpretation. From the IT perspective, it holds \textit{data}, not \textit{information} or \textit{knowledge}. We are building on the fact that any information that we would like to store within the time series data cube, even the provenance, can be represented by some value and can be treated as an axis from the data point of view. 

The \texttt{Time Series Cube} class as seen on Fig. \ref{fig:model} is the main class of \textit{Time Series Cube DM}. It extends the \texttt{Sparse Cube} class taken from \textit{N-Dimensional Cube DM}. This means that we are reusing what was already defined within \texttt{Sparse Cube} class coming from \textit{Sparse Cube DM} and add new parameters in the \texttt{Time Series Cube} class.

The UML representation of our model can be seen on Fig. \ref{fig:model}. On this figure we describe two data models, i.e., \textit{Dataset DM} and \textit{Cube DM}. The \textit{Axis Domain DMs} package is collecting external models that we reference. \textit{Time Series Cube DM} imports classes from the \textit{DataSet DM} model. 

A time series data cube can bee distinguished from a more generic sparse data cube by having the \texttt{dataProductType} attribute with value \texttt{"timeseries"}. 

The \textit{DataSet DM}, \textit{Cube DM} and \textit{Axis Domain} packages seen on Fig. \ref{fig:model} are described below.

\subsection {DataSet DM}
Everything you see in the \textit{Dataset DM} package on Fig. \ref{fig:model} is imported, i.e., not changed within the scope of this document. 

The \textit{DataSet DM} holds common metadata for all \texttt{Time Series Cube} class instances collected within this dataset and stores them in the \texttt{ObsDataset} class. Typically that would be \texttt{DataID} class attributes like \texttt{creator}, \texttt {Curation} class metadata like \texttt{publisherDID} or \texttt{releaseDate} and \texttt{Provenance} class metadata like \texttt{facility} or \texttt{instrument} identifying the origin of the data. A \texttt{Target} class can also have its metadata specified, if the dataset has a common observation target.

The \textit{SparseCube Dataset} class extends \textit{ObsDataset} class and puts restriction on what can be stored within this collection. An instance of the \textit{SparseCube Dataset} class has an attribute \texttt{sparseCubeData} which holds the collection of \textit{SparseCube} class instances, i.e. \texttt{Time Series Cube} instances.

\subsection {Cube DM}
In this section we speak about relevant classes from the \textit{Cube DM} package, i.e., \texttt{Sparse Cube} class and \texttt{Time Series Cube} class as described on Fig. \ref{fig:model}.

\subsubsection{Sparse Cube class}
On Fig. \ref{fig:model} we show the \texttt{Sparse Cube} class that we are re-using from \textit{N-Dimensional Cube DM}. The \texttt{Time Series Cube} class is extended from this one.

\subsubsection{Time Series Cube class}
A client can identify the \texttt{Time Series Cube} instance by having the keyword \texttt{timeseries} in the \texttt{dataProductType} attribute of its instance. By using only \textit{ColumnRef} class which is already referenced from \textit{Axis Domain models}, we override the \textit{Cube DM} dependency on axis domains, e.g., \textit{STC v2.0 DM}.

A client reading the data cube might want to know what should it treat as independent (coordinates) and dependent (data) values - they are collected within the \texttt{independent\_axes} and in \texttt{dependent\_axes} attributes of the \texttt{Time Series Cube} instance.  

\subsubsection{ColumnRef class}
 \texttt{independent\_axes} and in \texttt{dependent\_axes} inside \texttt{Time Series Cube class} use the \texttt{ColumnRef} which is a primitive element of the serialization used. In VOTable this would be \texttt{FieldRef} element, in FITS table just the \texttt{name} of the column.

If a specialized client wants to find more about specific metadata and data of an axis (its frame, domain, all related parameters and fields), it will find this information where he is used to it - in metadata of the dataset and in the parameters of the column itself. In VOTable this would be special \texttt{Param} elements at the beginning of the table and attributes of \texttt{Field} element, e.g., \texttt{name}, \texttt{type}, \texttt{unit}, \texttt{ucd}, etc.

And because this client is specialized towards that domain, it understands metadata in the referenced domain-specific data model (e.g., gravitational wave domain, spectral domain, etc.) and we don't need to make it part of \textit{Time Series Cube DM}.

\subsubsection{Quantity class}
This class was originally part of the \textit{Cube DM} but we moved it out because statistical information is useful for images and spectra, basically any kind of numerical data. We are still mentioning it here though because we believe that it will have bigger impact on \textit{Cube DM} than other types of data. 

Its main role is to describe statistical distribution of data within the axes of our data cube allowing us to make more informed decisions about what subset of the data we would like to filter. Knowing mean and sigma values or even quantiles of the data and filtering it based on these (e.g. "I want to filter only points that are 3-5 sigma from the mean value.") will provide much better results than trying to guess the ranges based only on minimum and maximum value of that axis.

The Quantity class is referencing the \textit{ColumnRef} class as it must say which column (also error column) it is describing in the dataset. Making it dependent only on this primitive type is again in accordance with the loose coupling principle we are using - making the higher level models independent of each other.

\subsection {Axis Domain DMs}
An answer to what all can be described with an axis (the axis domain) is not within the scope of this document. If we want to describe our spectral coordinates (e.g., list of filters) by the \textit{Photometry DM} or in \textit{STC DM} or add a completely new model for my axis domain (e.g., Gravitational Wave data), it does not affect the \textit{Time Series Cube DM}. 

These models, however, do always reference a primitive type of their serialization, such as \texttt{ColumnRef}. By using this primitive type dependency (loose coupling for all of our data models), we are essentially making \textit{Cube DM}, \textit{Quantity DM} and \textit{Axis Domain DMs} independent of each other.

\subsection{Time Series Cube DM summary}
With the growing amount of models and their diversity within the IVOA, keeping models independent of each other is the only viable approach to their evolution and maintenance. By only referencing a primitive type instead of an external model or even importing every detail of that model, the \textit{Time Series Cube DM} does not need to change every time an \textit{Axis Domain Model} changes or a new one is created - the abstraction of that domain will be still the same. This information is valid also for the \textit{Quantity DM}.


\section{Discovery - ObsCore}
\label{chap:discovery}
The ObsCore protocol is the natural option for discovery of our time series data cubes. In ObsCore 1.1, the most crucial parameters are added for the description of axis lengths for the basic types of axes, i.e., \texttt{s\_xel} for spatial axis length, \texttt{t\_xel} for temporal, etc., so the client knows what is the size of the cube it is about to download.  The usual implementation of ObsCore model discovery is using TAP protocol for access to the tables holding the time series data.

At current point of time, the ObsCore suffices our requirements for discovery and anyone can add new parameters to his ObsCore service - except the \textit{DataLink} (\cite{std:DataLink}) metadata discovery. This point is under discussion and, in the end, it may will not demand change to the ObsCore TAP standard.

\section{Serialization}
We decided to discuss the VOTable format as the first one to support as it is the easiest one to form discussions upon and afterwards close them with a deterministic XSD that can validate whether the serialization is strictly following the standard or not. This section focuses on serialization of the metadata and data defined within the \textit{Time Series Cube DM}, not on serialization of the data model structure - that will be handled by VO-DML.

\texttt{GROUP} XML element can be used as an equivalent of the data model classes described in Chap. \ref{chap:model}. These are linked to the IVOA data models by annotations. The \textit{Time Series Cube DM} itself then contains \texttt{independent\_axes} and \texttt{dependent\_axes} \texttt{GROUP} elements which hold the actual information about the axis. 

The metadata, i.e., common information for every point in that axis, is referenced in the group as \texttt{PARAMRef} elements and the data (actual measurements that have usually different values for every point of this axis) is referenced by \texttt{FIELDRef} elements. 

These \texttt{FIELDRef} elements point to \texttt{FIELD} elements, holding data of the VOTable. The order of these is the same as the order of data values stored in the standard \texttt{DATA} element of the VOTable (here we can find TR and TD elements of the VOTable). 

Axis Models are again serialized as \texttt{GROUP} elements and are referenced from the independent or dependent axis collections by the model using \texttt{GROUPRef}.

The basic example of how such a serialization would look like can be seen on Fig. \ref{fig:group_a_1} and more can be found in the Chap. \ref{chap:cases}.

Please note that we enhanced and completed the serialization examples for version 1.1 and updated them in \ref{chap:cases}.

\section{Supported science cases}
\label{chap:cases}

In this chapter we describe how the most typical time series use cases are supported. We show that our approach can support all of the usual use cases, such as simple light curves or groups of light curves, in the same way as completely generic time series data like provenance linkage or frequency-based analysis.

We focus mainly on the support of time series data other than light curves. We will describe the science cases collected and suggest a VOTable serialization example for each one of these examples, so we can see that the \textit{Time Series Cube DM} is easily serializable.

\subsection{Requirements}

The formalization of basic time series use cases comes from the \textit{CSPTimeseries} science cases document (\cite{TSUseCase}). In addition to these use cases, we collected a few others from LSST or LIGO requirements for time series data models. In this chapter, we take these science cases one by one and explain to implement them with \textit{Time Series Cube DM} . 

We took the first three logical groups of light curves from the \textit{CSPTimeseries} document (\cite{TSUseCase}) and added two more, the complex light curves and gravitational waves, as examples of what can be accomplished with the model.

\subsection{Science case definitions}
The science case definitions are defined from the requirements and are as follows.

\begin{itemize}
\item {Simple Light curves} - Combine photometry and light curves of a given object/list of objects in the same photometric band. We refer to the Group A in the \textit{CSPTimeseries} document mentioned above as \textit{simple light curves}.

\item {Groups of light curves} - Combine photometry and light curves of a given object/list of objects in different photometric bands. We refer to the Group B in the \textit{CSPTimeseries} document and slightly generalize it as \textit{groups of light curves}, i.e., not only collecting light curves for filters, but also holding other types of axes, e.g., polarization.

\item {Generic Time series} - Time series other than light curves. Basically anything what has at least one time-based axis can be put here. We refer to the Group C in the \textit{CSPTimeseries} document as \textit{generic time series}.

\item {Complex light curves} - Special requirements on light curve data: 
  \begin{itemize}
  \item{Provenance tracking}:  These use cases require to track original image or its cutouts for every point of the light curve, e.g., link to SIAP cutouts of the relevant region in the original image which can be opened in a graphical client. We are able to transfer the data and describe the metadata for this use case.
  \item {Probability distribution}: Another use case computes and transfers probability distribution functions for the time series, e.g., computed simple Gaussian distribution parameters for the coordinate errors from which the light curve was generated). We are able to transfer the data but for metadata description, there would have to be created a new kind of Probabilistic DM as seen on Fig. \ref{fig:model}. However, when such model comes, we just add the \texttt{model} reference to the \textit{CubeAxis} metadata, without any changes to the \textit{Time Series Cube DM} serialization.
  \item {\textit{DataLink} for time series}: A basic use case for \textit{Time Series Cube DM} combined with \textit{DataLink} would be storing data linkage for the whole time series instance, e.g., \textit{DataLink} to the periodogram of this time series. Another one would be filtering of light curve points on a supported criteria, e.g., filtering only high quality points of a light curve. Here we are struggling with the metadata transfer for the ObsCore TAP discovery described in Chap. \ref{chap:discovery}. The problem is that with \textit{ObsCore DM} we can get also other data than time series and the \textit{DataLink} does not support metadata restriction based on a \texttt{dataProductType}. We are able to construct such a \textit{DataLink} for the light curves, the only problem is with transferring the metadata about such a domain over ObsCore TAP service.
 
\end{itemize}  
 \item {Power spectra, Gravitational waves}: Special use case for generic time series data - we show here how to store not only simple time axis but also its derivatives, e.g., frequency. For Gravitational waves, this would be tracking for example ASD (strain per rtHz) against Frequency (Hz) as seen on the Gravitational wave tutorial (\cite{GravWave}).
\end{itemize}

\subsection{Simple Light curves}
This is the basic use case supported up to now with \textit{SimpleTimeSeries} (\cite{std:SimpleTimeSeries}). Now, we can easily support it in the \textit{Time Series Cube DM}. An example of the \textit{DataSet} metadata can be seen on Fig. \ref{fig:group_a_1}, the definition of the cube on Fig. \ref{fig:group_a_2} and \ref{fig:group_a_2_1} and  the first row on Fig. \ref{fig:group_a_3}. 

The light curve rendered within SPLAT-VO client ((\cite{paper:splat})) can be seen on Fig. \ref{fig:splat_ssa1} and \ref{fig:splat_ssa2}. These images were taken on real data coming within our reference implementation for server and client. More about the reference implementation can be found in Chap. \ref{chap:implementation}.

\subsubsection{Examples}
\begin{figure}[h!]
\includegraphics[width=1\textwidth]{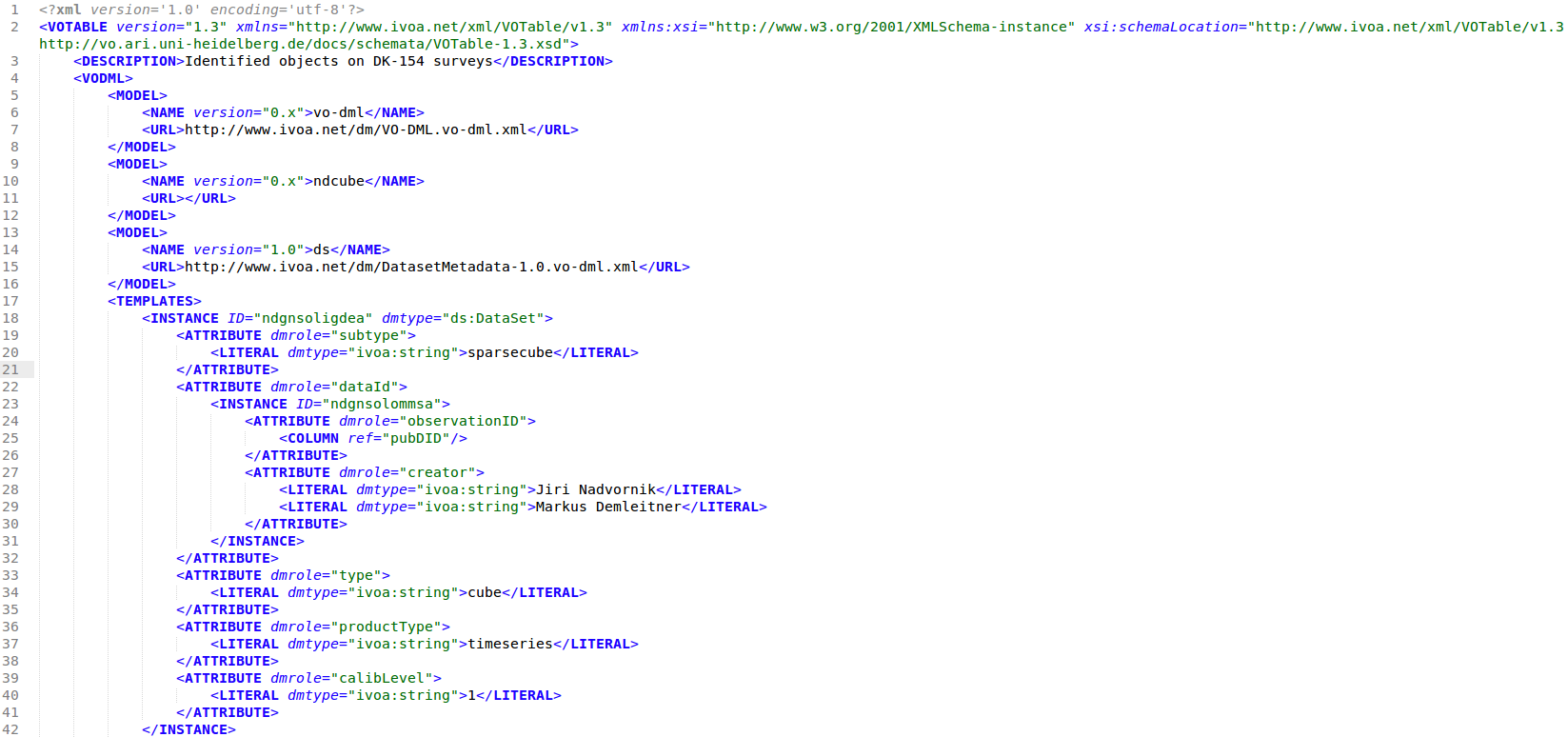}
\centering
\caption{Dataset metadata serialization via VO-DML. There is a dataset (collection) of sparse cubes. The dataset contains the metadata, e.g., the \texttt{observationID} or \texttt{creator} metadata. All of the serialized metadata are annotated against their respective counterparts in \textit{Dataset DM} and \textit{Time Series Cube DM} using the \texttt{dmRole} attribute.}
\label{fig:group_a_1}
\end{figure}

\begin{figure}[h!]
\includegraphics[width=1\textwidth]{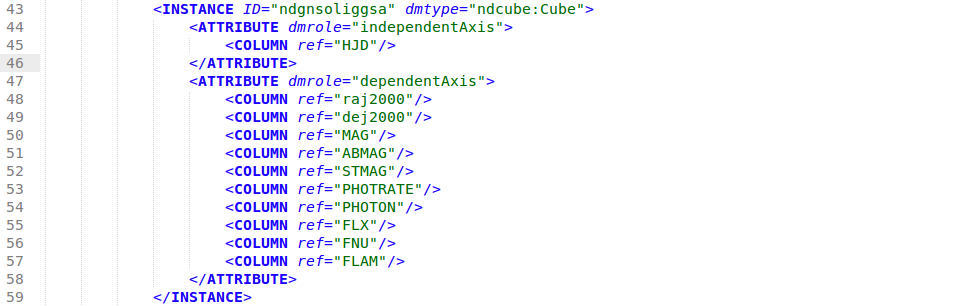}
\centering
\caption{Time series data cube metadata serialization. Here we can see the \texttt{ndcube:Cube} \texttt{dmtype} along with the axis definitions. The independent axes like temporal or spatial are referencing the \texttt{FIELD} elements holding their data. It works the same way for both independent and dependent axes.}
\label{fig:group_a_2}
\end{figure}

\begin{figure}[h!]
\includegraphics[width=1\textwidth]{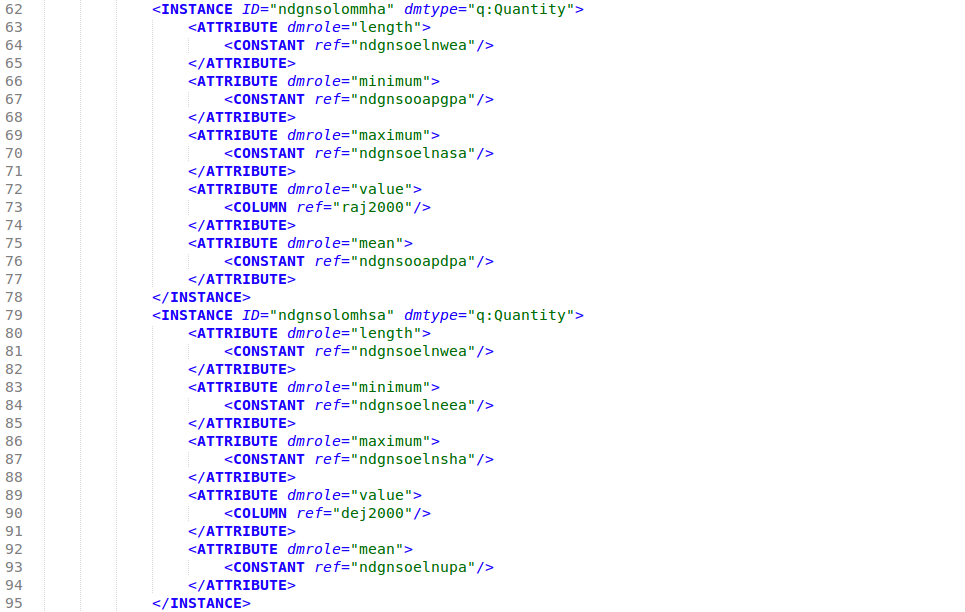}
\centering
\caption{Statistical information about the data cube. In here we can see the \texttt{q:Quantity} \texttt{dmtype}. The statistical information is held within the \texttt{constant} elements and the reference to the column that is actually holding the data can be recognized by \texttt{dmrole="value"} attribute. Note that for describing statistical information we don't need to have the context held in axis domain metadata.}
\label{fig:group_a_2_1}
\end{figure}

\begin{figure}[h!]
\includegraphics[width=1\textwidth]{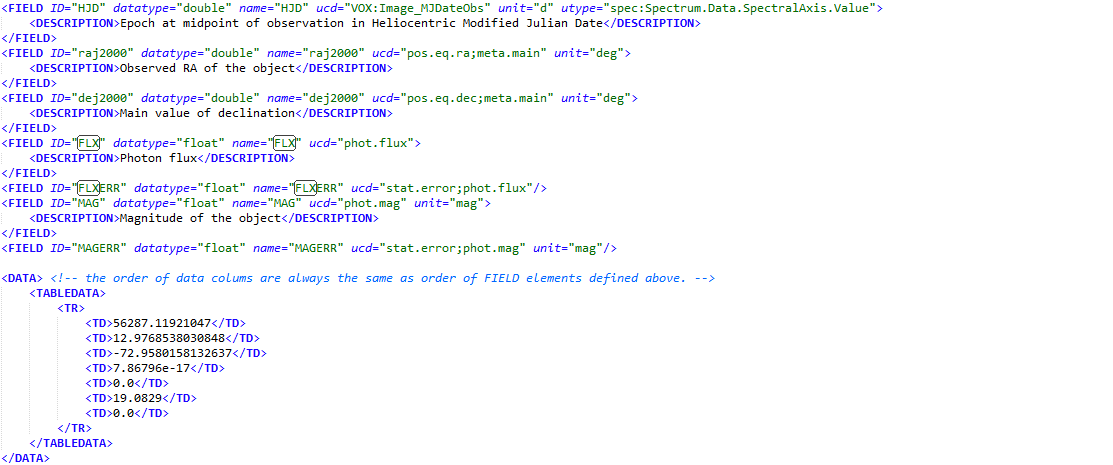}
\centering
\caption{Time series data cube data serialization. Here we can see the \texttt{FIELD} elements referenced from the time series cube and quantity metadata. This is a basic light curve holding the heliocentric Julian date for temporal axis, equatorial coordinates and two dependent axes with measurements - flux and magnitude.}
\label{fig:group_a_3}
\end{figure}

\begin{figure}[h!]
\includegraphics[width=1\textwidth]{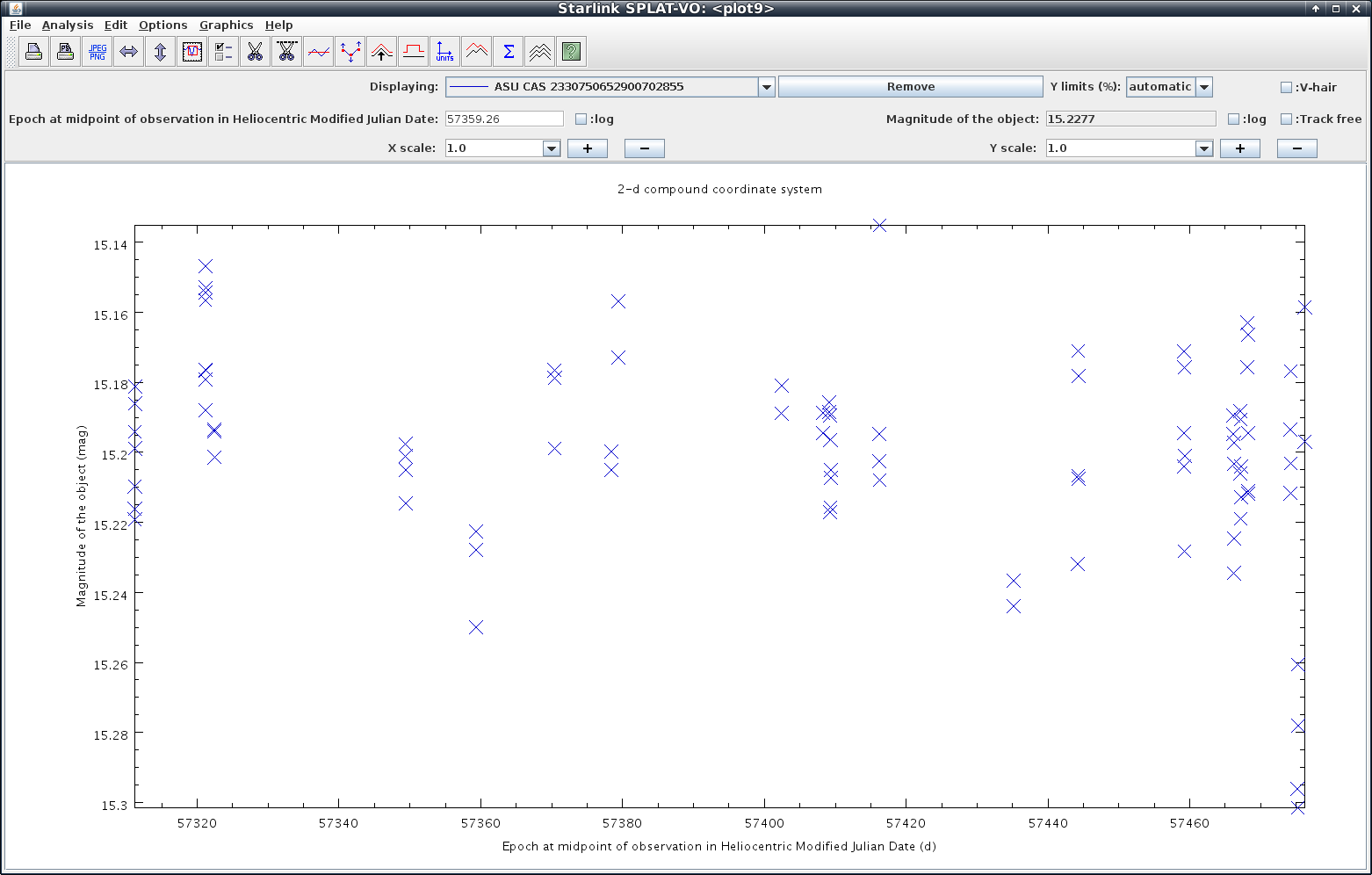}
\centering
\caption{Light curve within SPLAT-VO. The light curve is rendered from VOTable coming within time series data cube format. Note that SPLAT-VO is able to distinguish the magnitude and inverts the axis, as well as the units for the temporal axis. These are coming from the axis domain metadata.}
\label{fig:splat_ssa1}
\end{figure}

The examples of simple light curves can be seen on Fig. \ref{fig:group_a_1}, Fig. \ref{fig:group_a_2}, Fig. \ref{fig:group_a_2_1} and Fig. \ref{fig:group_a_3}. The plot rendered within SPLAT-VO can be seen on Fig. \ref{fig:splat_ssa1}.

\subsection{Groups of light curves}

The use case of storing light curves for one object in multiple photometric bands can be supported simply by addition of one independent axis - the spectral axis. Having the light curve points in multiple bands, e.g., UBVRI, is nothing more than having one more discrete spectral axis with coordinates that can have 5 distinct values. If the spectral axis is described in continuous intervals, the error \texttt{FIELD} would be meaningful, for just a list of filter names we are storing only data value elements, i.e. \texttt{value} attribute of  \texttt{Quantity}, with textual information.

It is evident that adding more such "groups" (other independent axes) will produce a lot of duplicated values in the \texttt{DATA} element of our VOTable. However, this is just a question of optimization.

If the need arises, for start, compressing algorithms such as EXI (\cite{std:EXI}) can be used for immediate value of eliminating the duplicities. A cleaner approach would be a table normalization during the transfer - the independent axes (coordinates) would be stored in separate \texttt{TABLE} elements (the same goes for FITS) that would be referenced by the dependent axes records. This, however, would need an extension to VOTable standard to allow referencing not only between elements but also within data. The requirement is that a value in one data column must be able to reference another value within another table.

The optimization of serialization is not the main aim of this document but can be expanded in the future if deemed necessary.

\subsubsection{Examples}
\begin{figure}[h!]
\includegraphics[width=1\textwidth]{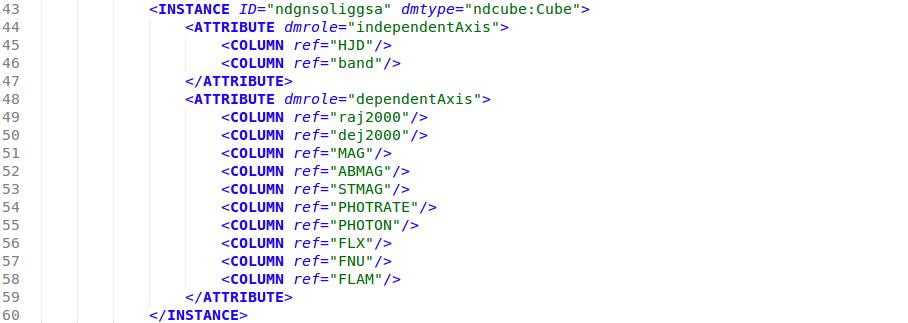}
\centering
\caption{Groups of light curves time series data cube metadata serialization. The \textit{Dataset} and \textit{Quantity} information are the same, the only change we need to do is to add one more axis to the independent collection. This means that the dependent axis (actual measurements, i.e., magnitude, flux) can hold values for any combination of the date and spectral axis values. }
\label{fig:group_b_1}
\end{figure}

\begin{figure}[h!]
\includegraphics[width=1\textwidth]{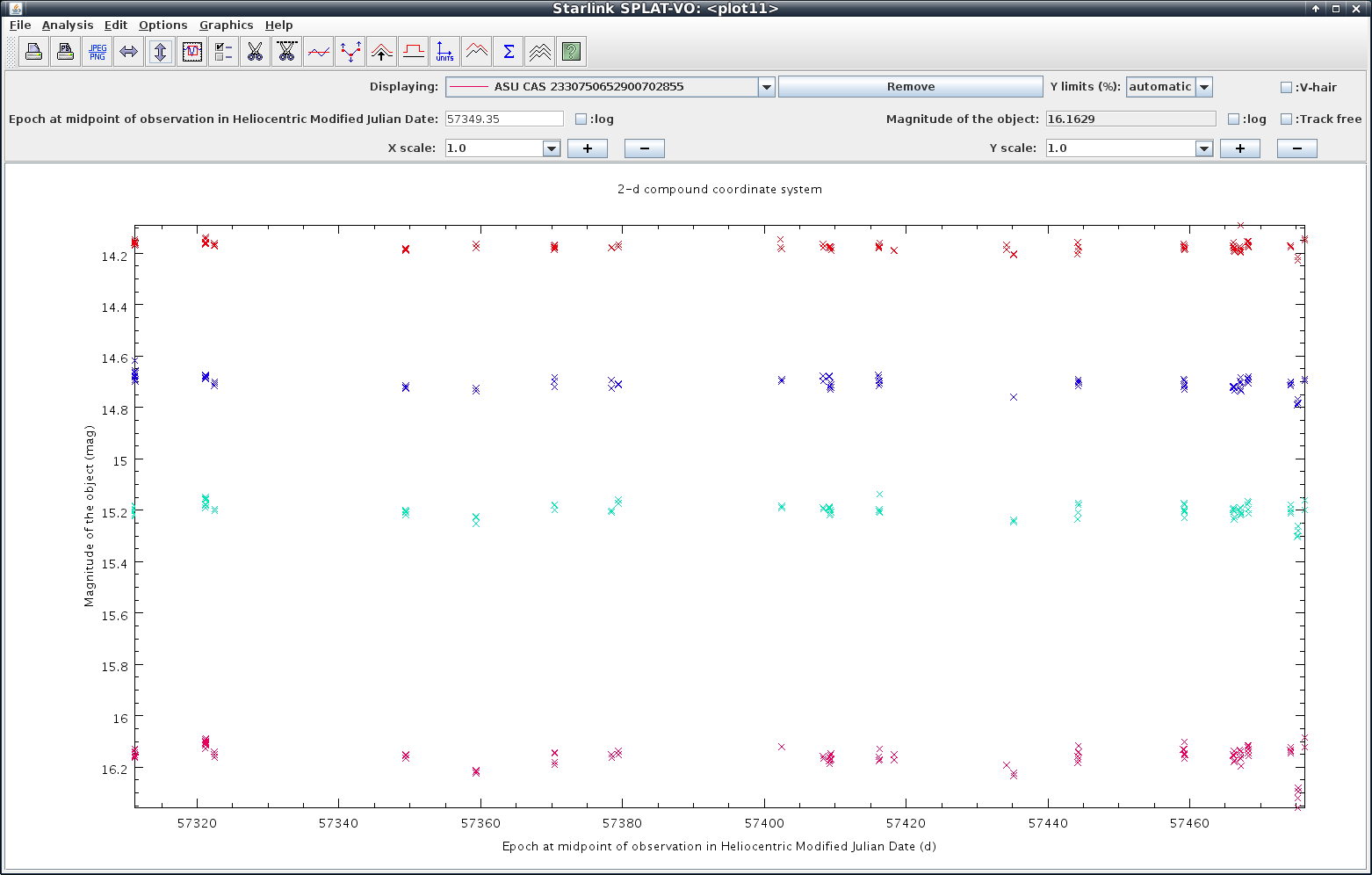}
\centering
\caption{Group of light curves within SPLAT-VO. As we are still using a valid VOTable serialization, our SPLAT-VO client can render the time series data cube for a light curve taken in 4 filters without modification. The spectral axis is the 3rd rendered axis and is expressed in colors instead of axis coordinates. The axis labels are taken from axis domain metadata already modeled within Photometry DM}
\label{fig:splat_ssa2}
\end{figure}

The XML example for groups of light curves can be found on Fig. \ref{fig:group_b_1}. The plot rendered within SPLAT-VO can be seen on Fig. \ref{fig:splat_ssa2}.

\subsection{Generic time series}
 \label{chap:group_c}
The Generic time series are the main motivation for using \textit{Time Series Cube DM}, as any kind of data that can be structured into series can be represented by a data cube. In here we also show that we can describe our metadata by using \texttt{PARAM} elements which our client can understand by parsing \texttt{UCDs}. Or, we can describe those by using external data models that the \textit{Time Series Cube DM} only references. 

The fact that the \textit{Time Series Cube DM} will not know about the details of every domain metadata is crucial as it enables us to describe the metadata with our own parameters if a standard model is not yet available, without the need of including them in \textit{Time Series Cube DM}. After the model is standardized, e.g., \textit{STC v2.0, Provenance DM, Probability DM}, etc., we just need to add new stuff defined within that model without the need to change anything within the serialization defined by \textit{Quantity DM} or \textit{Cube DM}.

The axes derived from the basic types can be still tracked as the same type. The frequency for a power spectrum is still a time series in its own sense and we can easily support it within \textit{Time Series Cube DM}. It is true that the the frequency is not technically a sparse axis that we chose as a basis for the time series data cubes, but logically a power spectrum or periodogram can be understood as time series. Moreover, the \textit{Sparse Cube DM} places a restriction on \textit{Time Series Cube DM} to support sparse axes, not to restrict itself only to them.

Another option for supporting these cases would be simply to move our generic definitions from \textit{Time Series Cube DM} to \textit{Cube DM} and not worry about whether a periodogram is a time series when we already can agree that i definitely can be described as a data cube.

\subsubsection{Examples}
\begin{figure}[h!]
\includegraphics[width=1\textwidth]{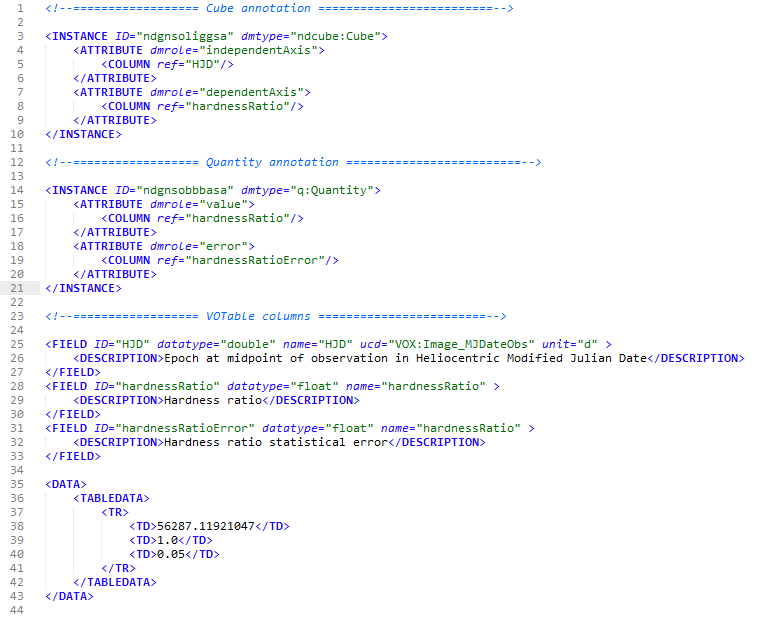}
\centering
\caption{Hardness ratio time series example. The data cube enlists the axes. The errors for these axes can be found in the \textit{Quantity} metadata (along with the statistics, which we omit here for simplicity) and the column definitions which are storing the actual data. When a model for hardness ratio arrives, it would certainly specify ucd for the hardnessRatio \texttt{FIELD} element.}
\label{fig:group_c_1}
\end{figure}

A basic example of a generic time series can be seen on Fig. \ref{fig:group_c_1}. Focus on the highlighted model attribute of the hardness ratio dependent axis. Even if the axis domain model does not exist yet, we still want to describe the structure of the data (values and errors of this axis) and store the rest in \texttt{PARAM} elements, which are not part of the data-perspective view provided by \textit{Time Series Cube DM}. When the axis domain model for hardness ration arrives, we just add new elements from it to the VOTable without the need to change any existing, maintaining full backwards-compatibility.

\subsection{Complex light curves}
This group is already describing science cases gathered apart the ones described in \textit{CSPTimeseries} document. We have collected these requirements from the most organizations most pressed by the needs for a standard, easily extendable solution.

\subsubsection{Examples}
\begin{figure}[h!]
\includegraphics[width=1\textwidth]{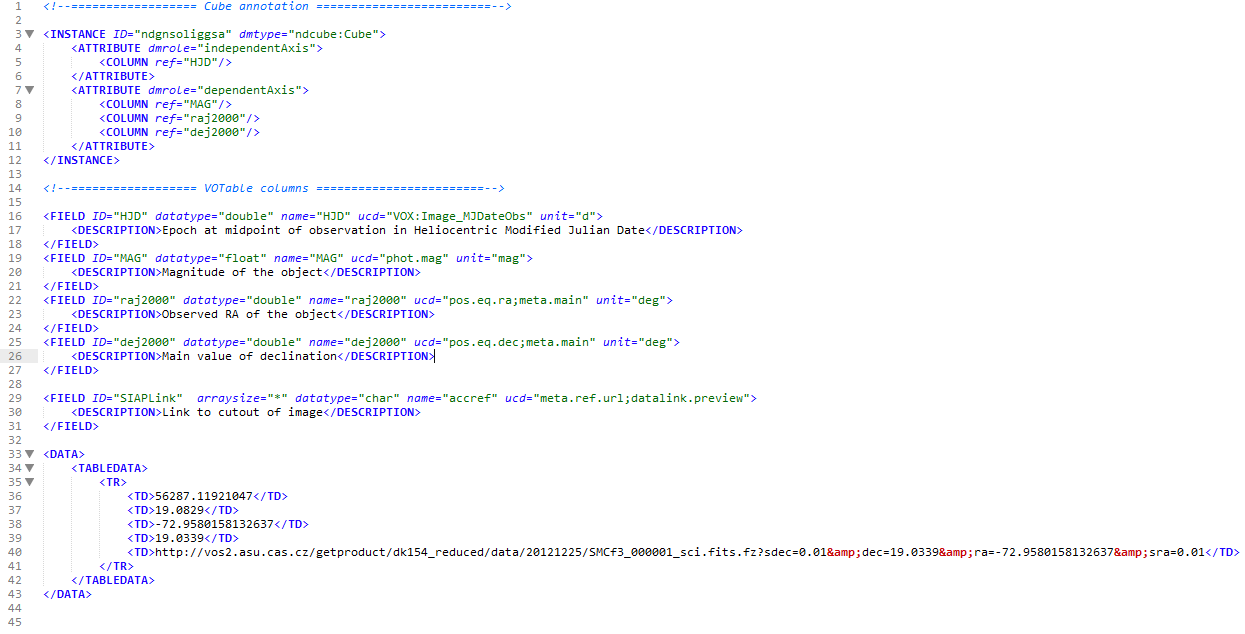}
\centering
\caption{Linkage towards original image cutout for one point of the light curve. The data can be seen in the last column of the table with cutout parameters taken from the right ascension and declination of this point of the curve. Note that while the cube could mention the \texttt{SIAPLiink} column as one of its axes, any statistical distribution or plotting of its values would not be meaningful. That is the reason why it is not part of the cube and is described entirely by the \texttt{FIELD} element that might be forged by the \textit{Provenance DM} in the future. }
\label{fig:lsst_1}
\end{figure}

One of the most important things when processing time series data is to be able to link the points of a light curve to the original image it was coming from. An outlier on the light curve could be a static noise or some other kind of error on the image but it could be also a valid burst in the object's magnitude. We can identify this on the first look if we have the link to the original data for each point of the light curve. The sample serialization of such provenance tracking can be seen on Fig. \ref{fig:lsst_1}.

\begin{figure}[h!]
\includegraphics[width=1\textwidth]{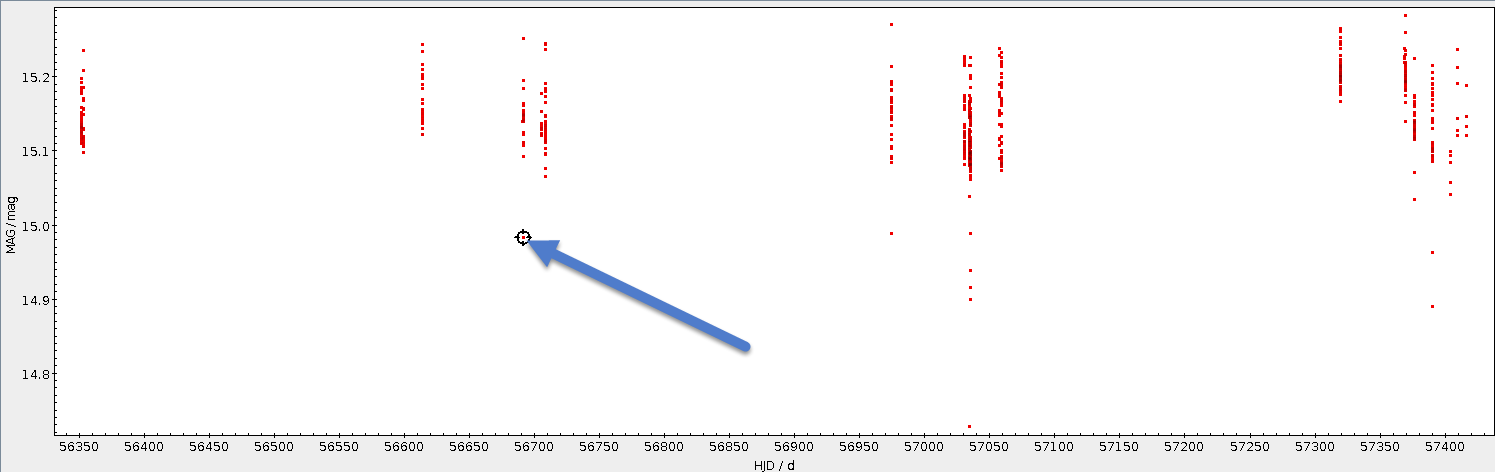}
\centering
\caption{A light curve rendered in Topcat client. The graph represents magnitude based on the HJD when the star was observed. After clicking on the highlighted point, Fig. \ref{fig:aladin_image} or Fig. \ref{fig:aladin_cutout} follows.}
\label{fig:topcat_lc}
\end{figure}

\begin{figure}[h!]
\includegraphics[width=1\textwidth]{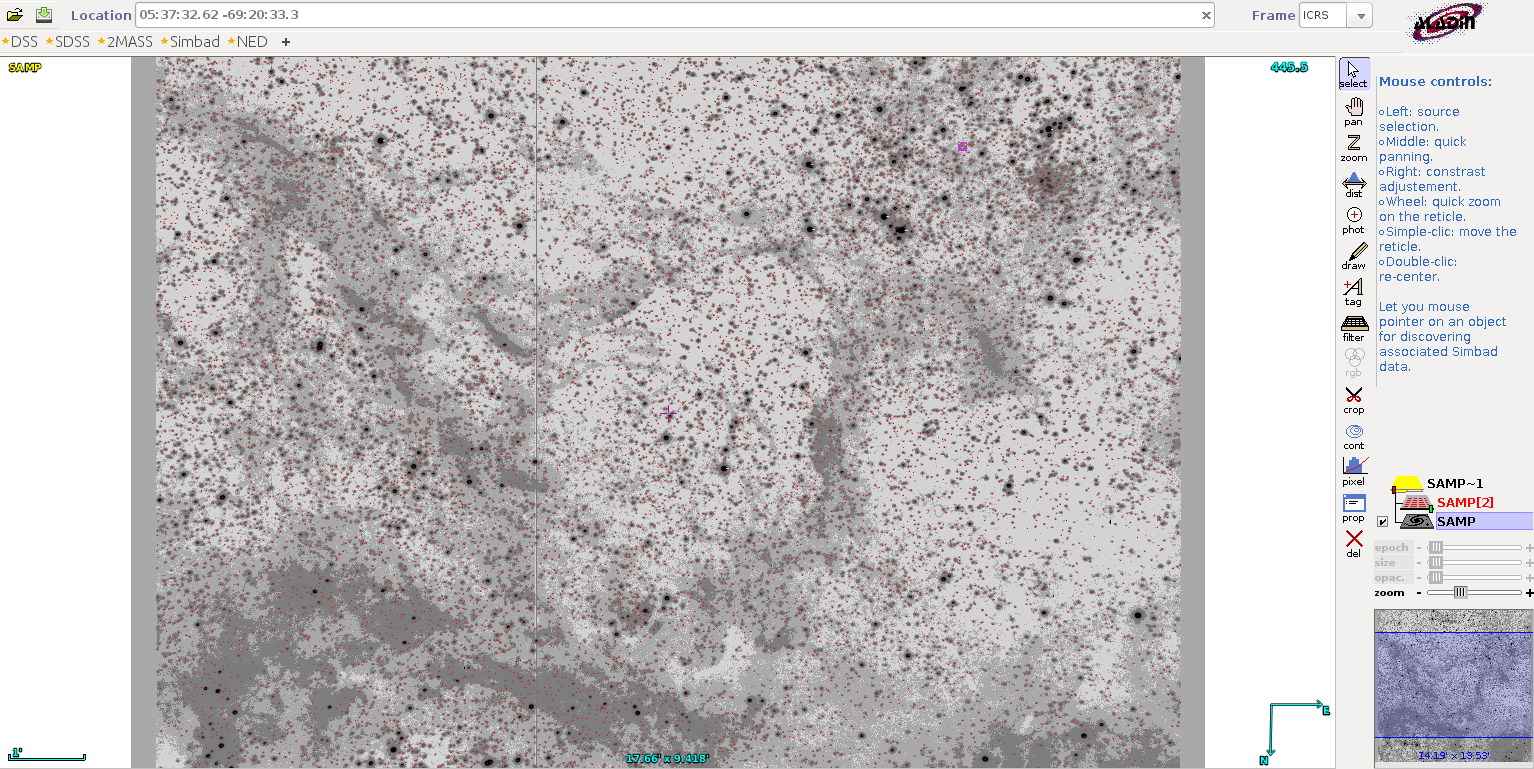}
\centering
\caption{Original image for one point of a light curve. After clicking on the point, the image from which it originates is displayed in Aladin.}
\label{fig:aladin_image}
\end{figure}

\begin{figure}[h!]
\includegraphics[width=1\textwidth]{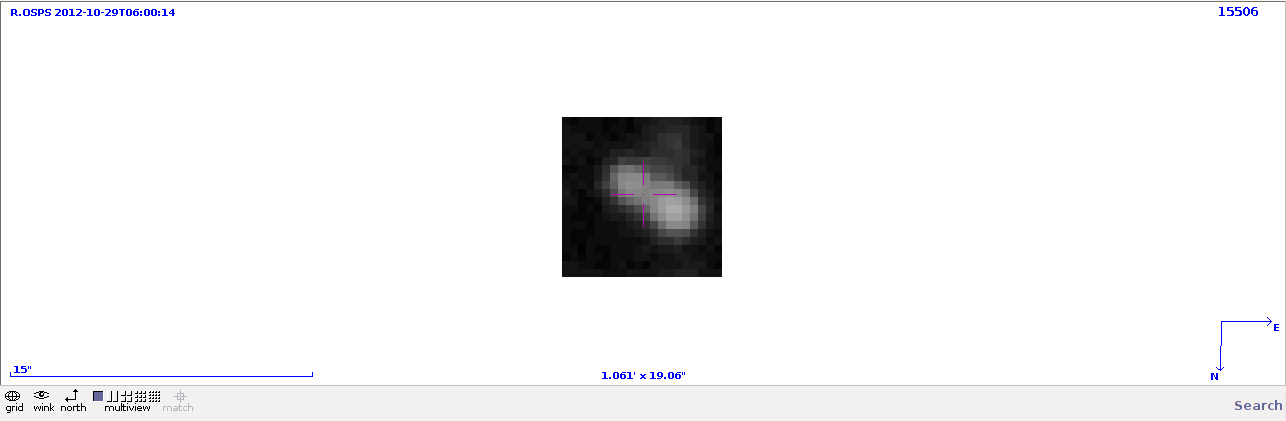}
\centering
\caption{Small cutout from original image for one point of a light curve. If a SIA cutout link is sent instead of the SIA link to the whole image, Aladin will render it the same way as the whole original image.}
\label{fig:aladin_cutout}
\end{figure}

The implementation on server side is easy and we plan this to be a very simple operation on the client side too. Just one click on the point of the light curve can send the coordinates via SAMP to Aladin client which displays the original image. The light curve plotted within TopCat client can be seen on Fig. \ref{fig:topcat_lc}. With the metadata \textit{Time Series Cube DM} provides, Topcat could automatically identify what to do when user clicks on a point of the light curve. In our case, we want to take the URL of the original image from which this point of the curve originates and send it to Aladin via SAMP, see Fig. \ref{fig:aladin_image}. We are also able to send just a cutout instead of the whole image link, as seen on Fig. \ref{fig:aladin_cutout}. 


\subsection {Gravitational waves}
\label{chap:grav_wave}
This example is quite similar to the one described in Chap. \ref{chap:group_c}. The only difference is a temporal axis describing a frequency (as time derivative) and then another dependent axis that contains measured data for gravitational wave analysis. This use case is taken from the LIGO tutorial site (\cite{GravWave}).

\begin{figure}[h!]
\includegraphics[width=0.8\textwidth]{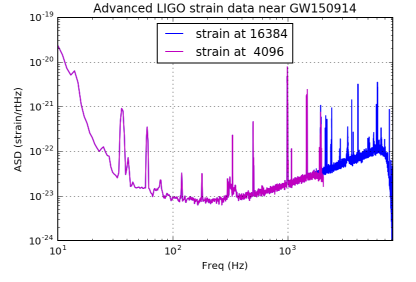}
\centering
\caption {LIGO time series plotting example. This graph is showing strain for frequency between \(10^{1}\) Hz and \(10^{4}\) Hz for two different samplings of 4096 Hz and 16384 Hz which are expressed in colors.}
\label{fig:ligo_splat_1}
\end{figure}

In our specific uses case, we describe a time series data cube storing gravitational wave analysis, as can be seen on Fig. \ref{fig:ligo_splat_1}. We have a temporal axis of \texttt{frequency} rendered as \texttt{x axis} and the \texttt{strain/rtHz} dependent axis rendered as \texttt{y axis}. The second independent axis holding sampling rate frequency is rendered as 3rd dimension in colors.

Storing the sampling rate frequency as an independent axis would make sense if it would be a configuration of the instrument, theoretically providing different values at the same point of time when measuring on different frequencies. Otherwise its just another dependent axis (and a very primitive one, just a multiple of the values for different sampling frequencies).

We assume that we have the more complicated option here that different sampling frequencies can provide different measurements.

\subsubsection{Examples}
\begin{figure}[h!]
\includegraphics[width=1\textwidth]{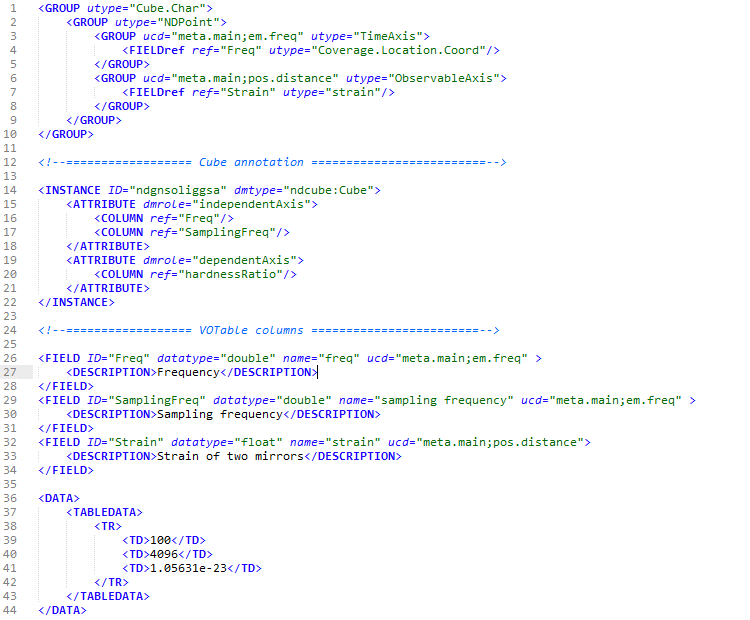}
\centering
\caption{VOTable example for gravitational Wave data. The independent axes are here the frequency and sampling frequency (explained in Chap. \ref{chap:grav_wave}. The sample row is showing a strain for 100 Hz frequency sampled with 4096 Hz.}
\label{fig:ligo_1}
\end{figure}

The sample serialization and one row can be seen on Fig. \ref{fig:ligo_1}.

\section{Reference implementation}
\label{chap:implementation}
We are working on GAVO DaCHS implementation for server-side service providing time series data cubes in the \textit{Time Series Cube DM} format. The service is ready and providing real data, but we have only a development environment ready right now and is under constant change. If you want to have a look on the final version, you can find it on the following web page (\cite{vos2}).

We are also working on the SPLAT-VO modifications so it can understand the \textit{Time Series Cube DM} format. Since we are using standard VOTables for serialization of our data cubes, the modifications are more related to SPLAT-VO understanding the axis domain models like magnitude axis inversion or time axis unit identification. In the current version of the serialization, SPLAT-VO understands the data perfectly.

\section{Summary}
This is still a work in progress and will need a lot of discussion and polishing to make it to an IVOA recommendation. However, we believe that this document creates a solid foundation upon which we can continue building.

	
\bibliography{cubeDM}

\end{document}